\documentclass{iau}
\usepackage{graphicx}
\usepackage{bm}

\title[Radiative transfer in 3D MHD models of the solar chromosphere] 
{Three-dimensional simulations of scattering polarization and the Hanle effect in MHD chromospheric models}

\author[\v{S}t\v{e}p\'an]   
{J. \v{S}t\v{e}p\'an
}

\affiliation{Astronomical Institute of the Czech Academy of Sciences, Fri\v{c}ova 298, 251\,65 Ond\v{r}ejov, 
Czech Republic\\ email: {\tt jiri.stepan@asu.cas.cz}} 

\pubyear{2015}
\volume{305}  
\jname{Polarimetry: From the Sun to Stars and Stellar Environments}
\editors{K.N. Nagendra, S. Bagnulo, \\ R. Centeno, \& M. Mart\'inez Gonz\'alez, eds.}
\begin{document}

\maketitle

\begin{abstract}
Scattering line polarization and the Hanle effect are among the most important mechanisms for diagnostics of the solar and stellar atmospheres. The fact that real stellar atmospheres are horizontally inhomogeneous makes the spectral synthesis and interpretation very challenging because the effect of thermodynamic fluctuations on spectral line polarization is entangled with the action of magnetic fields. This applies to the spatially resolved as well as to the averaged spectra. The necessary step towards the interpretation of such spectra is to study the line formation in sufficiently realistic 3D MHD models and comparison of the synthetic spectra with observations. This paper gives an overview of recent progress in the field of 3D NLTE synthesis of polarized spectral lines resulting from investigations with the radiative transfer code PORTA.
\end{abstract}

\firstsection 

\section{Introduction}
\label{stepan-sec-intro}

The solar atmosphere is an inherently three-dimensional (3D) medium. In order to decipher its structure and dynamics we need to understand the implications of the spatial plasma inhomogeneity for the polarized line spectra \cite[(see Manso Sainz \& Trujillo Bueno 2011, for a basic investigation assuming small horizontal inhomogeneities)]{stepan-msjtb11}. Departures from rotational symmetry of physical interactions (the so-called symmetry breaking effects) leave their fingerprints in the emergent Stokes spectra and leave us with the problem of how to disentangle and quantify their relative contributions. With the exception of relatively simple cases, it can be a difficult task because the symmetry of the system can be broken in number of ways simultaneously only providing ambiguous information on the physical conditions.

In the theory of NLTE line formation one attempts to take into account as much 
of physical ingredients involved in the line formation as possible. Not surprisingly, the very problem of forward synthesis then becomes quite challenging. Even though approximative 1D models can provide a first glimpse of physical properties of spatially constrained objects such as solar prominences, flares, and Ellerman bombs, a full-3D study of the lines with non-negligible optical thickness is necessary. This is due to the fact that the 1D plane-parallel approximation fails if lateral optical thickness of the particular structure is not much larger than the photon thermalization length. In the case of strong spectral lines, the 1D approximation fails in all the above structures.

Not only the spatially constrained objects listed above call for use of spectropolarimetry of strong NLTE lines. Large portions of the solar chromosphere, especially of its quiet upper layers and the chromosphere-corona transition region, can only be studied via spectropolarimetry of such spectral lines in infrared, visible, and UV ranges. Such lines are being formed all over the solar disk and carry a valuable information on the thermal structure and magnetization of their ensuing regions of formation. Given the fact that interpretation of such spectra is highly non-trivial, our knowledge of physical conditions in the upper chromosphere remains much scarcer than in the case of the underlying photosphere.

The existing one-dimensional semi-empirical models, regardless of their practical usefulness, have difficulties reproducing simultaneously the intensity and polarization of multiple spectral lines. Part of this problem is due to the fundamental fact that the non-linear line formation problem in an inhomogeneous 3D atmosphere cannot be reproduced by a 1D model. As the solar chromosphere is highly inhomogeneous, it is not surprising that its full understanding demands line formation modeling in realistic 3D models.

This paper provides an overview of the NLTE modeling we have carried out applying the radiation transfer code PORTA \cite[(see \v{S}t\v{e}p\'an \& Trujillo Bueno 2013]{stepan13}, using as model atmosphere a 3D snapshot taken from a state-of-the-art radiative-MHD simulation of the solar atmosphere. Here we focus on the formation problem of spectral lines of the mid-to-upper chromosphere and of the transition region, paying particular attention to understand their scattering polarization.

\section{The problem of symmetry breaking}
\label{stepan-sb}

\begin{figure}[t]
\begin{center}
\begin{tabular}{ccc}
\includegraphics[height=4cm]{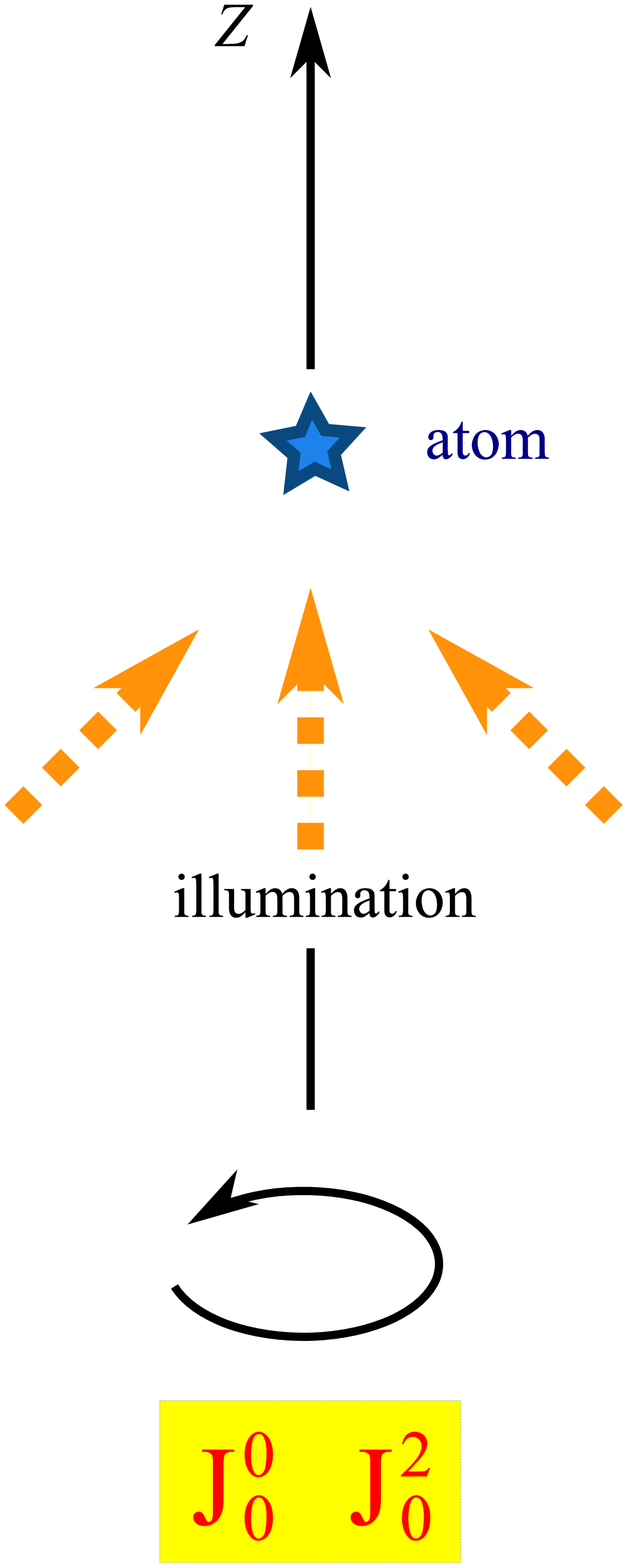} \hspace*{1cm} &
\includegraphics[height=4cm]{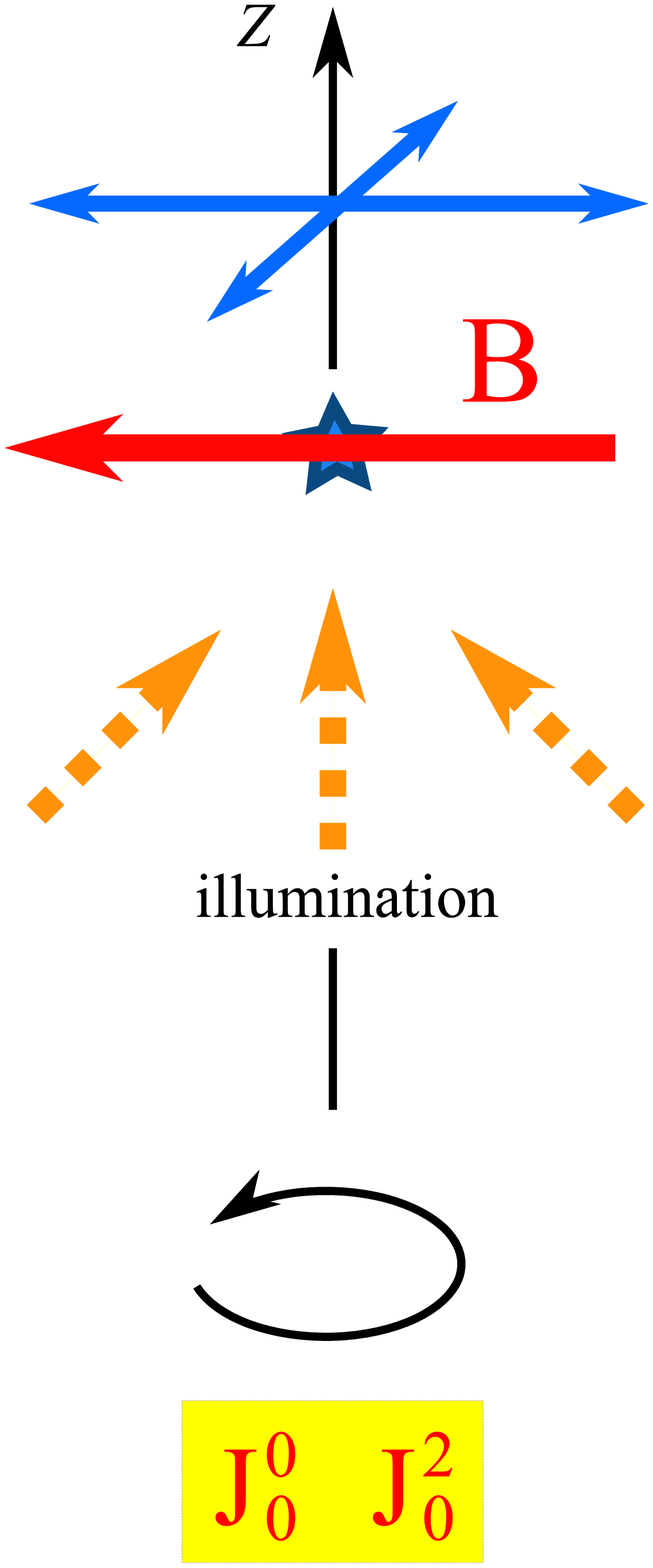} \hspace*{1cm} &
\includegraphics[height=4cm]{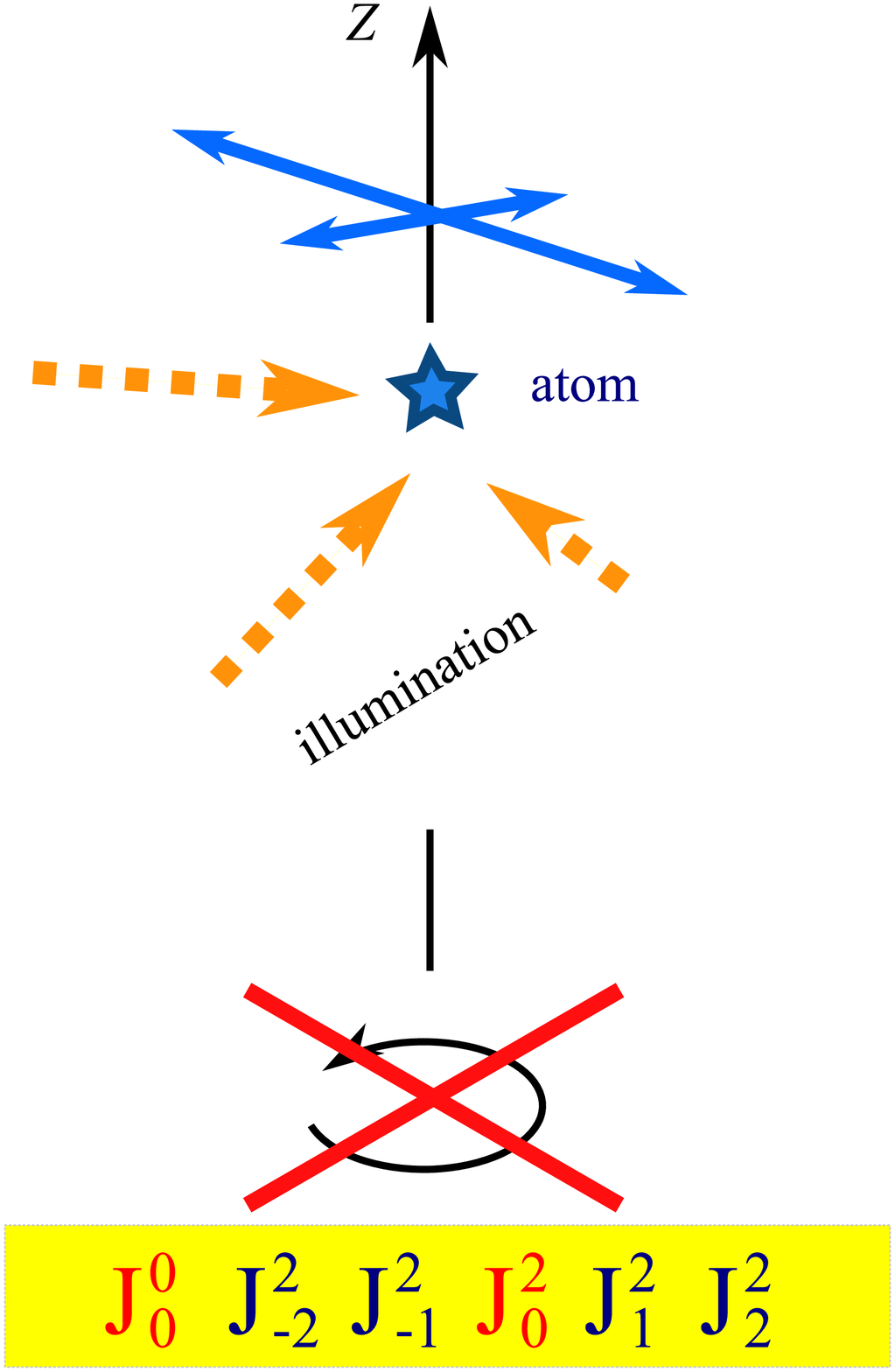}
\end{tabular}
\caption{
Breaking of axial symmetry of the radiation scattering problem. Left panel: In case of cylindrically symmetric illumination the linear polarization of forward-scattered radiation is zero. Central panel: In the presence of an inclined magnetic filed the axial symmetry of the problem is broken and the emergent radiation will generally be polarized.
Right panel: Similar result can be found in the absence of magnetic fields if the illumination is not cylindrically symmetric. Without additional information on the radiation field tensors $J^K_Q$ it may be impossible to distinguish the cases in the central and the right panel.
}
\label{stepan-fig-sb}
\end{center}
\end{figure}

Polarization of radiation can be viewed as a manifestation of a departure from the rotational symmetry of a light-emitting system. At the atomic level, the symmetry of the light scattering processes can be broken by presence of magnetic fields (Zeeman effect), electric fields (Stark effect), anisotropic velocity distribution of particles (impact polarization), anisotropy and polarization of the incident radiation (scattering polarization).

Scattering polarization is of particular interest for diagnostics of the solar atmosphere where the anisotropy of radiation gives rise to linearly polarized spectrum that is observed near the solar limb and that is known as the Second Solar Spectrum \cite[(Stenflo \& Keller  1997)]{stepan-sk97}. If we could approximate the solar atmosphere by a plane-parallel model which is unmagnetized, then the radiation field would be axially symmetric with respect to the local vertical and fully describable by the two radiation field tensor components, $J^0_0$ and $J^2_0$, that correspond to the mean intensity and alignment of the field, respectively \cite[(see Landi Degl'Innocenti \& Landolfi 2004 for more details)]{stepan-ll04}. These quantities are uniquelly determined by the stratification of the thermal structure of the model atmosphere.
In such a case there would be zero emergent polarization observable in the center of the solar disk due to the cylindrical symmetry of the problem (see the left panel of Fig.~\ref{stepan-fig-sb}).

The Hanle effect, i.e., modification of the quantum coherences among the Zeeman sublevels of an atomic level due to action of magnetic field followed by emission of polarized photon, allows us to infer information on the magnetic field.
As a matter of fact, to measure magnetic field via the Hanle effect we need to know a priori the nine independent radiation field tensorial components $J^K_Q$ in the region of spectral line formation.

\begin{figure}[t]
\begin{center}
\includegraphics[height=4cm]{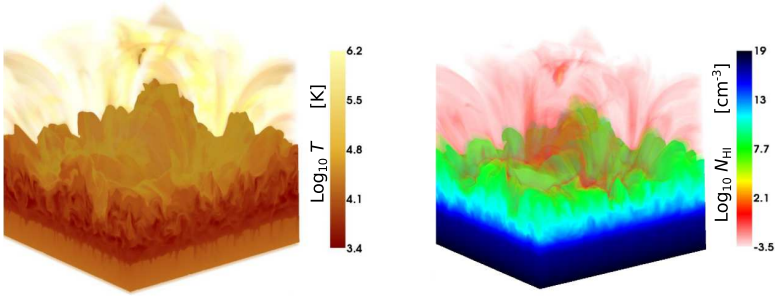}
\caption{
Visualization of the structure of the snapshot of the 3D MHD simulation. Left panel: Temperature. Right panel: Neutral hydrogen density. The spatial extension of the simulation is about $33\times 33$ arcseconds.
}
\label{stepan-fig-mhd}
\end{center}
\end{figure}

The forward-scattering Hanle effect \cite[(Trujillo Bueno 2001)]{stepan-jtb01} has been shown to be of great diagnostic interest for studying the magnetic field of coronal filaments through spectropolarimetry in the He\,{\sc i}\,10830\,\AA\ triplet \cite[(see Trujillo Bueno et al. 2002)]{stepan-jtb02}. Coronal filaments above quiet regions are illuminated by a spectrally flat radiation field coming from the underlying photosphere, which typically dominates the excitation of the filament helium atoms. The situation may be more complicated for optically-thick spectral lines that form in the solar chromosphere. High-resolution observations of some strong chromospheric lines actually show a strong variability of linear polarization signals of spectral lines sensitive to scattering polarization at the solar disk center \cite[(e.g., Bianda et al. 2011)]{stepan-bianda11}. If we restrict our discussion to the lines of the mid-to-upper solar chromosphere whose linear polarization is practically insensitive to the action of Zeeman effect, then one may be tempted to interpret the observed disk-center linear polarization patterns by the presence of magnetic fields that break the cylindrical symmetry of the line formation via the Hanle effect (see the central panel of Fig.~\ref{stepan-fig-sb}). However, to this end, a reliable model atmosphere of the observed region would be needed. Given an observation and a model atmosphere, one can solve the NLTE problem for possible magnetic field configurations and to constrain the solution \cite[(e.g., Manso Sainz \& Trujillo Bueno 2010, Anusha et al. 2011, Ishikawa et al. 2014)]{stepan-msjtb10,stepan-anusha11,stepan-ishikawa14}.

There are two main drawbacks in the above approach. Firstly, one needs to choose a particular semi-empirical model by some criteria. This problem can be solved to some extent because the choice can be done using quantitative criteria based on comparison of the line intensity profiles. More seriously, this approach cannot account for the effect of breaking of the cylindrical symmetry due to horizontal inhomogeneities of the atmosphere (see Tich\'y et al. 2015 in these proceedings).

Since the observations show that not only polarization but also intensity of spectral lines fluctuates from point to point on the solar disk, it follows that the thermal structure of the atmosphere is considerably inhomogeneous and these inhomogeneities can lead to additional symmetry breaking that may interfere with the action of the Hanle effect (see the right panel of Fig.~\ref{stepan-fig-sb}). To disentangle these two mechanisms may be a challenging task.

\section{Line synthesis in 3D MHD models}
\label{stepan-mhd}

\begin{figure}[t]
\begin{center}
\includegraphics[height=12cm]{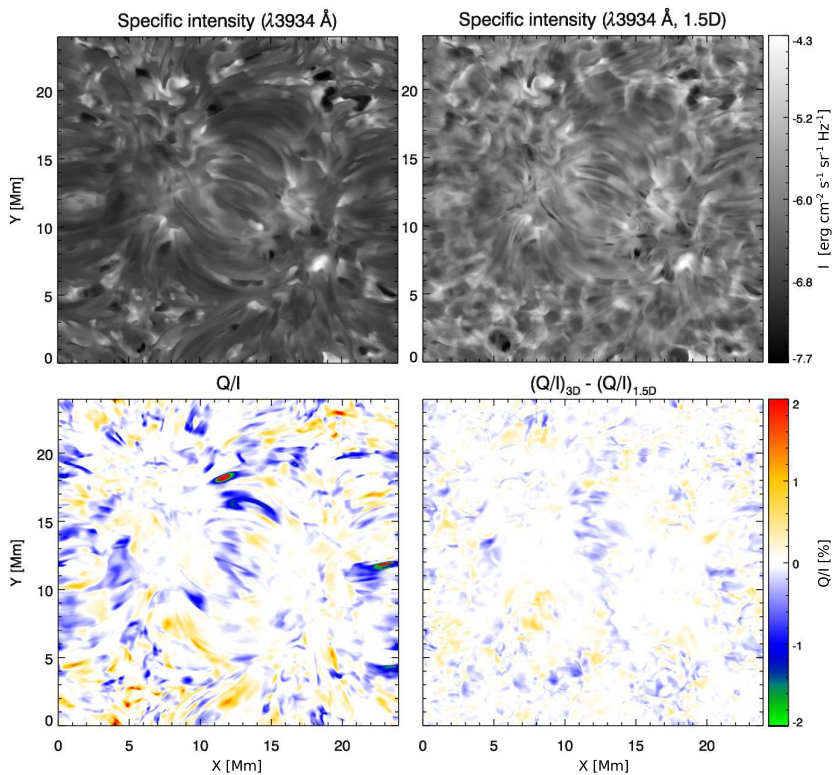}
\caption{
Emergent radiation at the center of the Ca\,{\sc ii}~K line. Top left: Line intensity in the full 3D model. Top right: Line intensity calculated in the 1.5D approximation. Bottom left: Fractional polarization $Q/I$ in the 3D solution. Bottom right: Difference of the $Q/I$ signal in the 3D and 1.5D solution. From \v{S}t\v{e}p\'an \& Trujillo Bueno (2015; in preparation).
}
\label{stepan-fig-K}
\end{center}
\end{figure}

\begin{figure}[t]
\begin{center}
\includegraphics[width=12.5cm]{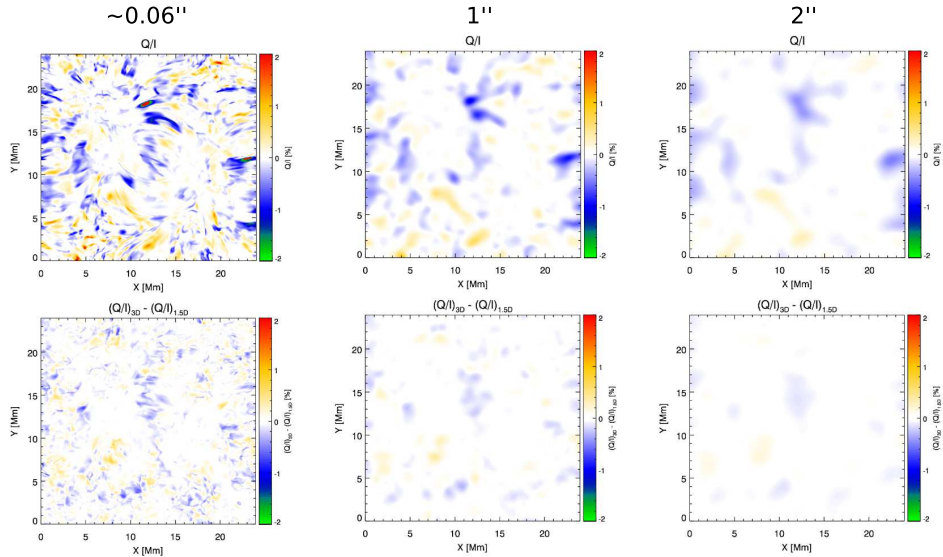}
\caption{
The effect of reduced spatial resolution. Top panels: emergent line-center $Q/I$ of the Ca\,{\sc ii}\,K line for three different spatial resolutions (from left to right: $0.06''$, $1''$, $2''$). Bottom panels: the difference between the emergent signals in 3D and 1.5D solutions for the same spatial resolutions. The results corresponding to the lower spatial resolution have been obtained by convolving the original polarization map with a 2D gaussian filter of the indicated full width at the half maximum. From \v{S}t\v{e}p\'an \& Trujillo Bueno (2015; in preparation).
}
\label{stepan-fig-resol}
\end{center}
\end{figure}

In order to investigate quantitatively the role of breaking of the cylindrical symmetry in the line formation, we have used a 3D model atmosphere which is snapshot 385 of the radiation MHD simulation ``en024048-hion" performed with the {\em Bifrost} code \cite[(Gudiksen et al. 2011)]{stepan-gudiksen11}, which takes into account non-equilibrium hydrogen ionization \cite[(for more details see Carlsson et al. 2015)]{stepan-carlsson15}.\footnote{This snapshot has recently been made publicly available through the Interface Region Imaging Spectrograph project \cite[(IRIS; De Pontieu et al. 2014)]{stepan-depontieu14} at the Hinode Science Data Center Europe (http://sdc.uio.no; see IRIS Technical Note 33)}. This 3D model atmosphere encompasses the upper part of the convection zone, the photosphere, chromosphere, transition region and corona (see Fig.~\ref{stepan-fig-mhd}).

We have used our 3D NLTE solver PORTA \cite[(\v{S}t\v{e}p\'an \& Trujillo Bueno 2013)]{stepan13} to make self-consistent synthesis of number of chromospheric and transition-region spectral lines of different species, namely the H \& K lines of Ca\,{\sc ii}, hydrogen Ly$\alpha$, and Mg\,{\sc ii}\,k. In all these strong chromospheric lines we have found that symmetry breaking due to horizontal structuring of the atmosphere plays an essential role.

\subsection{Impact of horizontal radiative transfer}

In Fig.~\ref{stepan-fig-K} we show a comparison of the emergent signals at the center of the Ca\,{\sc ii}\,K (3934\,\AA) line in the disk-center geometry calculated taking into account the horizontal transfer effects (left panels) and assuming every column of the model to be a plane-parallel atmosphere (the so-called 1.5D approximation; right panels). There is a clear difference between the line intensities. The 3D solution produces, on average, slightly lower intensity and the contrast of the intensity map is lower with respect to the 1.5D solution. This is due to the ``smoothing effect'' of the horizontal transfer which tends to reduce the coupling of radiation with the local thermal structure of the atmosphere \cite[(for a similar effect in the intensity of the H$\alpha$ line see Leenaarts et al. 2012)]{stepan-leenaarts12}.

Even more important is the impact of the 3D transfer on emergent linear polarization of the line. Firstly, the amplitudes of fractional linear polarization are of the order of a percent across all the field of view. When compared with the 1.5D solution in which only the Hanle effect and velocity fields are capable of breaking the local cylindrical symmetry, we find out that the relative error of the $Q/I$ signal caused by neglecting the 3D effects, calculated as a spatial average of the quantity $|(Q/I)_{3D} - (Q/I)_{1.5D}|/|(Q/I)_{3D}|$, is about 0.5. Vaguely speaking, the symmetry breaking is responsible for about 50\% of the emergent polarization signal in the center of the Ca\,{\sc ii}\,K line.

\begin{figure}[t]
\begin{center}
\includegraphics[width=12.5cm]{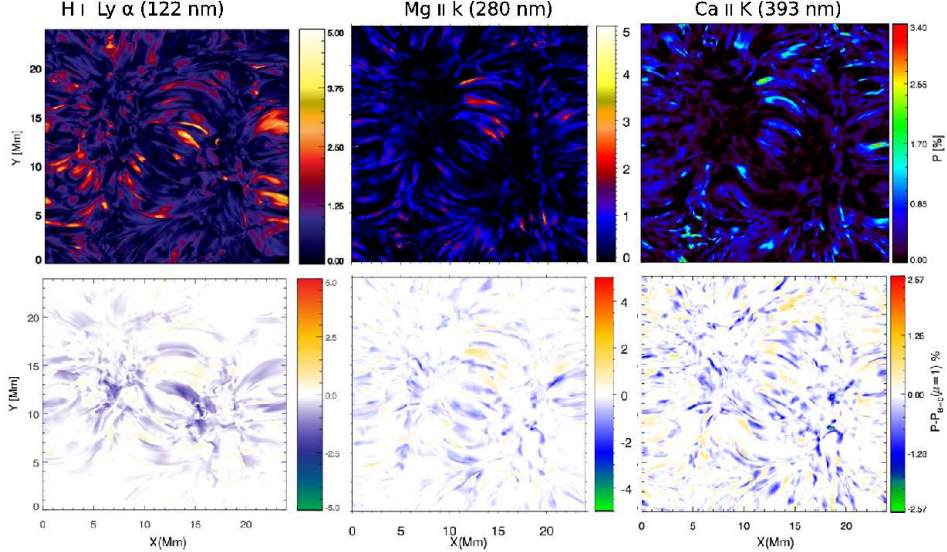}
\caption{
Top panels: Total linear polarization degree $P$ of three spectral lines. Left panel: Hydrogen Ly$\alpha$ \cite[(see \v{S}t\v{e}p\'an et al. 2015)]{stepan15}. Central panel: Mg\,{\sc ii}\,k (see del Pino Alem\'an et al. 2015; in preparation).  
Right panel: Ca\,{\sc ii}\,K (see \v{S}t\v{e}p\'an \& Trujillo Bueno 2015; in preparation). Bottom panels: Difference of the polarization amplitude $P$ shown in the upper panels and the total polarization degree in the model with zero magnetic field, $P_{B=0}$. All three spectral lines show predominating depolarization ($P-P_{B=0}<0$) due to the forward Hanle effect.
}
\label{stepan-fig-depolar}
\end{center}
\end{figure}

\subsection{The role of spatial resolution}

It is of interest to ask whether the role of symmetry breaking, which is due to small-scale perturbations of the atmosphere, can be reduced if we degrade the spatial resolution of the observations. This is indeed the case for the line intensity which converges to similar result as the 1.5D solution in the limit of no spatial resolution.

The situation is more complicated in the case of scattering polarization signals because the positive and negative $Q/I$ and $U/I$ values tend to cancel out if the spatial resolution is decreased (see Fig.~\ref{stepan-fig-resol}). Since the spatial variability of the magnetic field in this particular model is comparable to the spatial scale of the thermodynamic fluctuations, there is no global agent that would introduce a systematic preference of one sign of the Stokes parameters. Consequently, the quasi-random distribution of the polarization signals across the disk leads to a decrease of the average signals $Q/I$ and $U/I$ with the spatial resolution resolution element $R$ as $\propto R^{-1/2}$ and the same behavior is found for the difference between the 3D and 1.5D solution. In other words, even though the absolute difference between the 3D and 1.5D solution decreases with increasing $R$, the relative error of the signals remains practically independent of the resolution.

The above conclusion applies to the average polarization signals at the solar disk center. In some regions of the atmosphere in which the magnetic field is smooth on larger spatial scales, the effect of local symmetry breaking may in fact be reduced by using a partially spatially integrated observations. We also note that non-zero net polarization is found at larger heliocentric angles \cite[(see \v{S}t\v{e}p\'an et al. 2015)]{stepan15}.

\subsection{Sensitivity of the linear polarization signals to the magnetic field}
\label{stepan-infer}

Our recent results of 3D NLTE modeling of the hydrogen Ly$\alpha$ line indicate that magnetic field tends to depolarize the spatially averaged signals observed out of the solar disk center \cite[(\v{S}t\v{e}p\'an et al. 2015)]{stepan15}. In the spatially resolved case of the disk-center observation, the polarization in most of the surface points of the model is reduced rather than increased. This is an opposite conclusion to the traditional image of the forward Hanle effect based on the plane-parallel approximation in which the inclined magnetic field can only create polarization. We have found this conclusion to be valid for all the lines of the blue and UV part of the spectrum but The level of overall depolarization depends on the particular spectral line and its susceptibility to the symmetry breaking and to the Hanle effect (see Fig.~\ref{stepan-fig-depolar}).

\begin{figure}[t]
\begin{center}
\includegraphics[width=7cm]{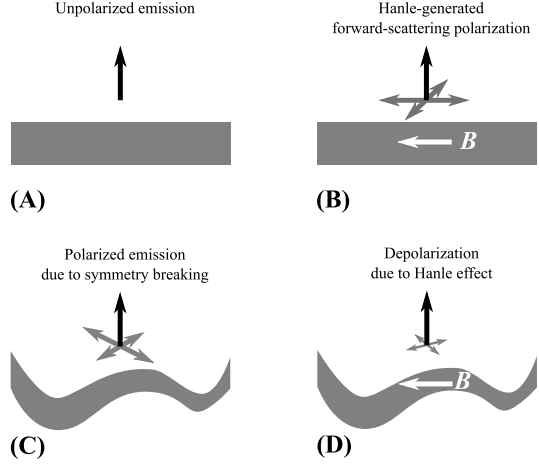}
\caption{
Qualitative explanation of depolarization of the UV/FUV lines by the Hanle effect in forward scattering. Panel A: in a 1D plane parallel unmagnetized atmosphere the disk-center radiation is unpolarized. Panel B: in the presence of an  inclined magnetic field disk-center polarization is created. Panel C: if the atmosphere is unmagnetized but horizontally inhomogeneous, the emergent radiation becomes polarized due to the ensuing symmetry breaking. Panel D: if a magnetic field is present in such a 3D atmosphere, and if the symmetry breaking due to the horizontal inhomogeneities is significant (i.e., $|J^2_{Q\neq 1}|$ components are of the same order of magnitude as $|J^2_0|$), then it is statistically more likely that the semi-randomly oriented $\vec B$ will depolarize instead of creating polarization.
}
\label{stepan-fig-cdep}
\end{center}
\end{figure}

Figure~\ref{stepan-fig-cdep} provides a qualitative explanation of the forward Hanle effect depolarization. In contrast to the forward scattering in a 1D plane parallel atmosphere, which is an example of a very peculiar symmetric configuration, magnetic field is more likely to cause depolarization in a sufficiently inhomogeneous medium. The ratio of the number of points with increased and decreased polarization depends on a particular distribution of magnetic fields and the relative importance of the symmetry breaking quantified by the $J^2_{Q\neq 0}$ components of the radiation field tensor. Not surprisingly, the depolarization occurs mainly in the points whose polarization is large already in the non-magnetized model, i.e., where the symmetry breaking is strong.

\section{Conclusions and future prospects}
\label{stepan-conc}

The aim of this paper was to point out the importance of the role of symmetry breaking effects in the strong chromospheric lines and the need to address these effects when interpreting the observations. The first step toward this goal is make forward synthesis of the lines of interest in sufficiently realistic model atmospheres. The second step must be to fully understand the phenomena involved in the 3D line formation process that are not existing in the simpler 1D models. In this article, we have pointed out few of our very recent findings and their consequences for the line formation. Other phenomena are still under investigation and some of them are still to be included in our modeling (such as the effects partial frequency redistribution that play an important role in formation of some lines).

The ultimate goal shall be to understand the magnetism and dynamics of the quiet chromosphere.
The spectropolarimetric forward synthesis in the 3D MHD models provides the tool by means of which these models can be contrasted with the observations and reliably tested (at least in the statistical sense). Possible inference of magnetic fields from comparison of the 3D models and observations is now the work in progress and we are pursuing several promising approaches. These approaches involve investigation of particular signatures in the line profiles that can be addressed to the action magnetic fields \cite[(see, for instance, \v{S}t\v{e}p\'an \& Trujillo Bueno 2010)]{stepan10}, study of the average atmospheric properties, and comparative analysis of multiple spectral lines with different sensitivities to the Hanle and Zeeman effects and to the symmetry breaking effects.

\begin{acknowledgments}
Financial support by the Grant Agency of the Czech Republic through grant \mbox{P209/12/P741} and project \mbox{RVO:67985815} is gratefully acknowledged.
I am indebted to Javier Trujillo Bueno and Tanaus\'u del Pino Alem\'an (IAC) for valuable 
discussions and ongoing collaborations on this research topic that made this work possible.
I am also grateful to Mats Carlsson (University of Oslo) and Jorrit Leenaarts (Stockholm University) for their support concerning the 3D MHD model.
The radiative transfer simulations presented in this paper were carried out with the MareNostrum supercomputer of the Barcelona Supercomputing Center (National Supercomputing Center, Barcelona, Spain). We gratefully acknowledge the computing grants, technical expertise and assistance provided by the Barcelona Supercomputing Center. We are also grateful to the European Union COST action MP1104 (Polarization as a Tool to Study the Solar System and Beyond) for financing short-term scientific missions at the IAC that facilitated the development of this research project.
\end{acknowledgments}


\end{document}